\documentstyle[12pt]{article}

\def\hybrid{\topmargin -20pt	\oddsidemargin 0pt
	\headheight 0pt	\headsep 0pt
	\textwidth 6.25in	
	\textheight 9.5in	
	\marginparwidth .875in
	\parskip 5pt plus 1pt	\jot = 1.5ex}
\def\cQ{{\cal Q}}
\def\cG{{\cal G}}
\def\cL{{\cal L}}
\def\cH{{\cal H}}
\def\ket#1{|{#1}\rangle}
\def\noi{\noindent}
\def\half{{1\over2}}
\def\baselinestretch{1.2}

\catcode`\@=11

\def\marginnote#1{}
\def\draftlabel#1{{\@bsphack\if@filesw {\let\thepage\relax
   \xdef\@gtempa{\write\@auxout{\string
      \newlabel{#1}{{\@currentlabel}{\thepage}}}}}\@gtempa
   \if@nobreak \ifvmode\nobreak\fi\fi\fi\@esphack}
	\gdef\@eqnlabel{#1}}
\def\@eqnlabel{}
\def\@vacuum{}
\def\draftmarginnote#1{\marginpar{\raggedright\scriptsize\tt#1}}

\def\draft{\oddsidemargin -.2truein
	\def\@oddfoot{\sl preliminary draft \hfil
	\rm\thepage\hfil\sl\today\quad\militarytime}
	\let\@evenfoot\@oddfoot	\overfullrule 3pt
	\let\label=\draftlabel
	\let\marginnote=\draftmarginnote
   \def\@eqnnum{(\theequation)\rlap{\kern\marginparsep\tt\@eqnlabel}%
\global\let\@eqnlabel\@vacuum}  }

\def\preprint{\twocolumn\sloppy\flushbottom\parindent 2em
	\leftmargini 2em\leftmarginv .5em\leftmarginvi .5em
	\oddsidemargin -.5in	\evensidemargin -.5in
	\columnsep .4in	\footheight 0pt
	\textwidth 10.in	\topmargin  -.4in
	\headheight 12pt \topskip .4in
88	\textheight 6.9in \footskip 0pt
	\def\@oddhead{\thepage\hfil\addtocounter{page}{1}\thepage}
	\let\@evenhead\@oddhead	\def\@oddfoot{}	\def\@evenfoot{} }

\def\numberbysection{\@addtoreset{equation}{section}
	\def\theequation{\thesection.\arabic{equation}}}

\def\underline#1{\relax\ifmmode\@@underline#1\else
	$\@@underline{\hbox{#1}}$\relax\fi}

\def\titlepage{\@restonecolfalse\if@twocolumn\@restonecoltrue
\onecolumn
     \else \newpage \fi \thispagestyle{empty}\c@page\z@
	\def\thefootnote{\fnsymbol{footnote}} }

\def\endtitlepage{\if@restonecol\twocolumn \else \newpage \fi
	\def\thefootnote{\arabic{footnote}}
	\setcounter{footnote}{0}}  

\catcode`@=12
\relax

\def\figcap{\section*{Figure Captions\markboth
	{FIGURECAPTIONS}{FIGURECAPTIONS}}\list
	{Figure \arabic{enumi}:\hfill}{\settowidth\labelwidth{Figure
999:}
	\leftmargin\labelwidth
	\advance\leftmargin\labelsep\usecounter{enumi}}}
 \relax
\def\tablecap{\section*{Table Captions\markboth
	{TABLECAPTIONS}{TABLECAPTIONS}}\list
	{Table \arabic{enumi}:\hfill}{\settowidth\labelwidth{Table
999:}
	\leftmargin\labelwidth
	\advance\leftmargin\labelsep\usecounter{enumi}}}
 \relax
\def\reflist{\section*{References\markboth
	{REFLIST}{REFLIST}}\list
	{[\arabic{enumi}]\hfill}{\settowidth\labelwidth{[999]}
	\leftmargin\labelwidth
	\advance\leftmargin\labelsep\usecounter{enumi}}}
 \relax
\makeatletter
\newcounter{pubctr}
\def\publist{\@ifnextchar[{\@publist}{\@@publist}}
\def\@publist[#1]{\list
	{[\arabic{pubctr}]\hfill}{\settowidth\labelwidth{[999]}
	\leftmargin\labelwidth
	\advance\leftmargin\labelsep
	\@nmbrlisttrue\def\@listctr{pubctr}
	\setcounter{pubctr}{#1}\addtocounter{pubctr}{-1}}}
\def\@@publist{\list
	{[\arabic{pubctr}]\hfill}{\settowidth\labelwidth{[999]}
	\leftmargin\labelwidth
	\advance\leftmargin\labelsep
	\@nmbrlisttrue\def\@listctr{pubctr}}}
 \relax
\makeatother
\newskip\humongous \humongous=0pt plus 1000pt minus 1000pt

\newif\ifdtup

\font\Scbig=cmss10 scaled\magstep1
\font\Scscr=cmss8 scaled\magstep1
\font\Scscrscr=cmss8
\newfam\Scfam
\textfont\Scfam=\Scbig
\scriptfont\Scfam=\Scscr
\scriptscriptfont\Scfam=\Scscrscr
\def\Sc{\fam\Scfam}

\relax
\hybrid
\def\lvm{\leavevmode\hbox to\parindent{\hfill}}

\def\thefootnote{\fnsymbol{footnote}}
\def\BE{\begin{equation}}
\def\EE{\end{equation}}
\def\BA{\begin{eqnarray}}
\def\EA{\end{eqnarray}}
\def\d{\partial}

\def\P{\Phi}

\def\tt{\bar\tau}

\def\lvm{\leavevmode\hbox to\parindent{\hfill}}
\def\bar{\overline}
\def\req#1{(\ref{#1})}

\def\L{\left}
\def\R{\right}

\def\sump{\sum_{p\geq-1}}

\def\BE{\begin{equation}}
\def\EE{\end{equation} \vskip 0.30\baselineskip}
\def\BA{\begin{array}}
\def\EA{\end{array}}

\def\noi{\noindent}

\def\frac#1#2{{\textstyle{{#1}\over{#2}}}}
\def\half{{1\over2}}

\def\Kr#1{\delta_{{#1},0}}
\def\ket#1{|{#1}\rangle}

\def\hL{\widehat{L}}

\def\cG{{\cal G}}
\def\cH{{\cal H}}

\def\cL{{\cal L}}
\def\cO{{\cal O}}
\def\cQ{{\cal Q}}

\def\open#1{\mbox{{\bf{#1}}}}

\def\oZ{{\open Z}}

\def\sJ{{\Sc J}}
\def\sL{{\Sc L}}

\def\ctop{{\Sc c}}

\def\Ups{\Upsilon}
\def\d{\partial}

\def\bc{background charge}

\def\cs{constraints}

\def\V{Virasoro}

\newif\ifold \oldtrue \def\new{\oldfalse}
\let\ssection=\section
\def\section{\setcounter{equation}{0}\ssection}

\begin{document}
\renewcommand{\theequation}{\thesection.\arabic{equation}}
\newcommand{\beq}{\begin{equation}}
\newcommand{\eeq}[1]{\label{#1}\end{equation}}
\newcommand{\ber}{\begin{eqnarray}}
\newcommand{\eer}[1]{\label{#1}\end{eqnarray}}
\begin{titlepage}
\begin{center}

\hfill IMAFF-94/6\\
\hfill hep-th/9411185\\
\vskip .5in

{\large \bf Topological Theories from \V\  Constraints on the  KP
 Hierarchy}
\vskip .8in

{\bf Beatriz Gato-Rivera and Jose Ignacio Rosado}\\
\vskip
 .3in

{\em Instituto de Matem\'aticas y F\'\i sica Fundamental, CSIC,\\ Serrano 123,
Madrid 28006, Spain} \footnote{e-mail addresses:
bgato, jirs @cc.csic.es}\\

\vskip .5in

\end{center}

\vskip .6in

\begin{center} {\bf ABSTRACT } \end{center}
\begin{quotation}
A conformal field theory can be recovered, via the Kontsevich-Miwa
 transform, as a solution to the \V\ \cs\ on the KP $\tau$ function.
 That theory, which we call KM CFT,
consists of $d \leq 1$  matter plus a scalar and a dressing
prescription: $\Delta=0$ for every primary field. By adding a spin-1 $bc$
system the KM CFT provides a realization of the $N=2$ twisted
topological algebra. The other twist of the corresponding untwisted $N=2$
superconformal theory is a DDK realization of the $N=2$
twisted topological algebra.

\end{quotation}
\vskip 1.0cm
\begin{center}
Talk given at the "28th International Symposium on the Theory
of Elementary Particles", Wendisch-Rietz (Germany), August 30 -
September 3, 1994.
\end{center}
\vskip 1.5cm

November 1994\\
\end{titlepage}
\vfill
\eject
\def\baselinestretch{1.2}
\baselineskip 16 pt
\section{Introduction}\lvm

To start let me review a couple of concepts about null states.
In general, a null state is a state with zero norm. In conformal
field theories (CFT's) null states are those states that are both
primary and secondary. That is, they are descendants built on primary
states by acting with the Virasoro generators (and possibly other
generators as well) and they also satisfy the highest weight
conditions that define primary states \cite{BPZ}.

In CFT's null states are orthogonal to any primary and any secondary
state (in the Hilbert space there may also exist states that are neither
primary nor secondary and null states need not be orthogonal
to them). Therefore when a null state enters a correlator of primary
and/or secondary fields, the correlator must vanish.

By using the commutation relation

\BE \L[L_n,\Psi(z)\R] = \biggl(z^{n+1}{\d\over\d z} +
 (n+1) \Delta z^n \biggr) \Psi(z) \EE

\noi
between the Virasoro generators and the primary fields (and other
analogous relations if there are other generators), the vanishing
of the correlator translates into a differential equation of the
same order as the level of the corresponding null vector. We will
refer to those differential operators as {\it decoupling operators}
since they express the decoupling of the null states

\BE \biggl({\rm decoupling\,\, operator}\biggr)
\biggl\langle \Psi(z_i)\prod_{j\neq i}
\Psi_j(z_j)\biggr\rangle=0 \,\,.\label{deceq}\EE

\noi
Here $\Psi(z_i)$ is singled out as the primary field on which the
null descendant was built.

\section{CFT's from Virasoro Constraints on the KP Hierarchy}\lvm

The \V\ \cs\ on the KP $\tau$ function $\ \sL_p\tau=0$,
$\ p\geq-1$, are \cs\
that are compatible with the KP flows and satisfy the \V\
algebra (half of it). They are given by

\BE\new\BA{rcl}\sL_{p>0} &=&\half\sum^{p-1}_{s=1}{\d^{2}\over\d
t_{p-s}\d t_s}+\sum_{s\geq 1}st_s {\d\over\d t_{p+s}}+\biggl(\sJ-
\half\biggr)(p+1){\d\over\d t_p}\\ \sL_0&=&\sum_{s\geq 1}st_s
{\d\over\d t_s}\\ {\Sc L}_{-1}&=&\sum_{s\geq 1}(s+1)t_{s+1}
{\d\over\d t_s} \EA\label{Vcs}\EE

\noi
Where $\sJ$ is the only degree of freedom.
They were written down independently by A.~Semikhatov
\cite{SeV} and P.~Grinevich and Yu.~Orlov \cite{GYO}, although the
final form with the degree of freedom $\sJ$ is due to the
first author.

Let us define the Kontsevich-Miwa (KM) transformation as

\BE t_r={1\over r}\sum_j n_j z^{-r}_j,\quad r\geq1\label{KMtr} \EE

\noi
where $n_j$ are real numbers and the coordinates $z_j$
belong to a set $\{z_j\}$ of points on the spectral complex plane
of the hierarchy. The KM transformation looks formally like
the Miwa transformation but it is not quite the same, since for Miwa
the parameters $n_j$ were essentially integer numbers \cite{Miwa}
\cite{DJKM}. He
was interested in expressing the bilinear Hirota equations as
finite-difference equations in the $n_j$. His attention
was on these parameters rather than on the points $z_j$. On the
other hand, more recently Kontsevich \cite{Kont} used the Miwa
transformation paying all his attention to the points $z_j$
(all the $n_j$ were set equal to a constant). Since for us the parameters
$n_j$ and the points $z_j$ are equally important, we call
\req{KMtr} the Kontsevich-Miwa transformation.

Intriguingly enough, using the KM transformation, a CFT can be
recovered as a solution to the \V\ \cs\ \req{Vcs}. Namely, for any
chosen $(z_i,n_i),\  z_i\in \{z_j\}$, and provided
 $\ 2\sJ-1={1\over n_i}-2n_i,\ $
the object
 \BE (\sump z_i^{-p-2} \sL_p) \label{zL}\EE

\noi
(sort of half the energy-momentum
tensor) KM-transforms into a decoupling operator \cite{SeKM}

\BE \L\{-{1\over 2n_i^2}{\d^2\over\d z_i^2}+{1\over n_i}
 \sum_{j\neq i}{1\over z_j-z_i}\L(n_j{\d\over\d z_i}-
n_i{\d\over\d z_j}\R)\R\} \,,\label{mastereq}\EE

\noi
corresponding to a level-2 null vector in a CFT of $d\leq1$
matter dressed by a scalar with zero \bc, so that $c=d+1$.
The Virasoro generators corresponding
to this CFT can be written, therefore, as

\BE L_n= L_n^{matter}-\half \sum_{m\in{\oZ}} :I_{n-m} I_m: \EE

\noi
The eigenvalues of $L_0$ and $I_0$ give the conformal weights
and $U(1)$ charges of the fields

\BE L_0\ket{\Psi_j} = \Delta_j \ket{\Psi_j}, \quad
    I_0\ket{\Psi_j} = n_j \ket{\Psi_j}\ .   \EE

\noi
Observe that the parameter $n_j$ plays the role of the $U(1)$
charge of the field $\Psi_j(z_j)$ sitting at the location $z_j$.
This theory, which we call KM CFT, has also a built-in dressing
prescription, $\Delta_j=-\half Q_M n_j$, that we call the KM
dressing for obvious reasons ($Q_M = \sqrt{1-d\over3}$ is the \bc\
of the matter).

The solution to the decoupling equation \req{deceq}
corresponding to the decoupling operator \req{mastereq}
 is therefore a correlator in which
 $\Psi(z_i)$ is a (2,1) or (1,2) dressed primary field with $U(1)$
charge $n_i$, and $\Psi_j(z_j)$ are fields with $U(1)$ charges $n_j$.

\vskip 0.2in

Now let us consider level $l$ decoupling operators in the KM CFT.
These KM-transform back into objects of the form \cite{BeSe3}

\BE \cO_l (\sL_p) (\sump z_i^{-p-2} \sL_p)  \EE

\noi
where $\cO_l$ is a differential operator, provided

\BE 2\sJ-1={l-1\over n_i}-{2 n_i\over l-1} \EE

For a given set of \V\ \cs\ (a given $\sJ$) this tunning
between the level $l$ and $n_i$ is a fortunate fact that ensures
that for a given $(z_i,n_i)$ there is at most one level at which
the KM transformation of the object \req{zL}
can take place.

One remark we should make is that the above is exactly true when
the decoupling operators correspond to null vectors built on $(1,l)$
or $(l,1)$ primary fields. The solution to the decoupling equation
\req{deceq} is therefore a correlator with
$\Psi(z_i)$ a $(l,1)$ or $(1,l)$ dressed primary field.

\section{KM and DDK Realizations of the Topological N=2 Twisted
Algebra.}\lvm

Let us concentrate on the KM CFT. By twisting the Virasoro
generators in the form

\BE \hL_n = L_n + \half Q_M (n+1)I_n \EE

\noi
we get a universal KM dressing: $\Delta_j=0$
 for every field $\Psi_j$.
This twisting modifies also the central charge and the scalar
\bc : $c=2$ and $Q_s=Q_M$, while previously $c=d+1$ and $Q_s=0$.
Now, by adding a spin-1 $bc$-system $(c_{gh}=-2)$ the KM CFT can
be embedded into a $N=2$ twisted topological theory \cite{BeSe2}
\cite{BeSe3}. The topological twisted $N=2$ algebra reads
\cite{EY} \cite{W-top}

\BE\new\BA{lclclcl}
\L[\cL_m,\cL_n\R]&=&(m-n)\cL_{m+n}\,,&\qquad&[\cH_m,\cH_n]&=
&{\ctop\over3}m\Kr{m+n}\,,\\
\L[\cL_m,\cG_n\R]&=&(m-n)\cG_{m+n}\,,&\qquad&[\cH_m,\cG_n]&=&\cG_{m+n}\,,
\\
\L[\cL_m,\cQ_n\R]&=&-n\cQ_{m+n}\,,&\qquad&[\cH_m,\cQ_n]&=&-\cQ_{m+n}\,,\\
\L[\cL_m,\cH_n\R]&=&\multicolumn{5}{l}{-n\cH_{m+n}+{\ctop\over6}(m^2+m)
\Kr{m+n}\,,}\\
\L\{\cG_m,\cQ_n\R\}&=&\multicolumn{5}{l}{2\cL_{m+n}-2n\cH_{m+n}+
{\ctop\over3}(m^2+m)\Kr{m+n}\,,}\EA\qquad m,~n\in\oZ\,.\label{topalgebra}
\EE

\noi
$\cL_m$ and $\cH_m$ are the bosonic generators corresponding to the
energy-momentum tensor and the topological $U(1)$ current
respectively, while $\cQ_m$ and $\cG_m$ are the fermionic generators
corresponding to the BRST current and spin-2 fermionic current.
The {\it topological central
charge} $\ctop$ is the true central charge of the $N=2$
superconformal algebra.

The KM realization of this algebra is

\BE\cL_m=\hL_m+l_m,\quad
l_m=\sum_{n\in{\oZ}}n\!:\!b_{m-n}c_n\!: \label{L}\EE

\BE\cH_m= - \sum_{n\in\oZ}:\!b_{m-n}c_n\!: +
{}\sqrt{{3-\ctop\over3}}I_m
\,, \quad  \cQ_m=b_m\,,\label{HQ}\EE

\BE\cG_m=2\sum_{p\in\oZ}c_{m-p}\hL_p+2{}\sqrt{{3-\ctop\over3}}
\sum_{p\in\oZ}(m-p)c_{m-p}I_p
{}+\sum_{p,r\in\oZ}(r-p):\!b_{m-p-r}c_rc_{p}\!:
{}+{}{\ctop\over3}(m^2+m)c_m\,,\label{G}\EE

\noi
and the chiral primary states split as $\ket\P = \ket\Psi \otimes
\ket0_{gh}$, where $\ket\Psi$ is a primary state of the KM CFT.

The null states of the KM CFT can be recovered from the topological
theory as follows \cite{BeSe2} \cite{BJI}.
 One constructs BRST-invariant ($\cQ_0$-invariant) topological null
states of bosonic type $\ket\Xi^{BQ}$ . Then, after the splitting of
the topological generators into their KM components, all ghost
contributions cancel out and one is left with

\BE \ket\Xi_{KM}^{BQ} = \ket\Ups \otimes \ket0_{gh}\ , \EE

\noi
where $\ket\Ups$ is a null state of the KM CFT.

\vskip .2in

Now it comes the question about the relatives of the KM topological
theory. Namely the {\it mother} $N=2$ untwisted superconformal theory
and the {\it sister} $N=2$ twisted topological theory corresponding
to the other possible twist of the untwisted superconformal
theory. After some computations one finds \cite{BeSe3}
 that the {\it mother} $N=2$
superconformal theory, satisfying the superconformal algebra
\cite{LVW}

\BE\new\BA{lclclcl}
\L[\cL_m,\cL_n\R]&=&(m-n)\cL_{m+n}+{\ctop\over12}(m^3-m)\Kr{m+n}
\,,&\qquad&[\cH_m,\cH_n]&=
&{\ctop\over3}m\Kr{m+n}\,,\\
\L[\cL_m,\cG_r^\pm
\R]&=&\L({m\over2}-r\R)\cG_{m+r}^\pm
\,,&\qquad&[\cH_m,\cG_r^\pm]&=&\pm\cG_{m+r}^\pm\,,
\\
\L[\cL_m,\cH_n\R]&=&{}-n\cH_{m+n}\\
\L\{\cG_r^-,\cG_s^+\R\}&=&\multicolumn{5}{l}{2\cL_{r+s}-(r-s)\cH_{r+s}+
{\ctop\over3}(r^2-\frac{1}{4})
\Kr{r+s}\,,}\EA\label{N2algebra}
\EE

\noi
is given by

\BE\cL_m=L_m+{1\over4}
(Q_L+Q_M)(m+1)I_m+l_m\,,\qquad
l_m=\sum_r\L(r+{m\over2}\R):\!b_{m-r}c_r\!:\,,\EE

\BE\cH_m=-\sum_r:\!b_{m-r}c_r\!: -\half(Q_L-Q_M)I_m\,,\qquad
 \cG^-_r=b_r\,,\EE

\BE\cG_r^+=2\sum_{n}c_{r-n}L_n
+\sum_{s,q}(s-q):\!b_{r-s-q}c_sc_q\!:
+\sum_{n}(Q_L n+(Q_M-Q_L)r
+\frac{1}{2}Q_M+\frac{1}{2}Q_L)
c_{r-n}I_n
{}+{}{\ctop\over3}(r^2-\frac{1}{4})c_r\,.\label{G32}\EE

\noi
where $Q_M$ and $Q_L$ are expressed in terms of $\ctop$ as

\BE Q_M=-{\ctop+3\over\sqrt{3(3-\ctop)}}\,,\qquad
Q_L={\ctop-9\over\sqrt{3(3-\ctop)}}\,.\EE

\noi
Here $\cG_r^\pm$ are spin-3/2 supersymmetric currents (the ghosts
are now a spin-3/2 $bc$ system).
 The twist that
gives rise to the KM topological theory is

\BE\new\BA{rclcrcl}
\cL^{(1)}_m&=&\multicolumn{5}{l}{\cL_m+\half(m+1)\cH_m\,,}\\
c^{(1)}_m&=&c_{m+\half}\,,&\qquad &b^{(1)}_m&=&b_{m-\half}\,,\\
\cH^{(1)}_m&=&\cH_m\,,&{}&{}&{}&{}\\
\cG^{(1)}_m&=&\cG_{m+\half}^+\,,&\qquad &\cQ_n^{(1)}&=&\cG^-_{n-\half}
\,,\EA\EE

\noi
while the other twist

\BE\new\BA{rclcrcl}
\cL^{(2)}_m&=&\multicolumn{5}{l}{\cL_m-\half(m+1)\cH_m\,,}\\
c^{(2)}_m&=&c_{m-\half}\,,&\qquad &b^{(2)}_m&=&b_{m+\half}\,,\\
\cH^{(2)}_m&=&-\cH_m\,,&{}&{}&{}&{}\\
\cG^{(2)}_m&=&\cG_{m+\half}^-\,,&{}&
\cQ_n^{(2)}&=&\cG^+_{n-\half}\,,\\
\EA\EE

\noi
provides a DDK realization \cite{DDK} of the topological algebra.
 That is, it consists of
minimal matter dressed by the Liouville ($\Delta=1$ for all the
primary fields, and $Q_s=\sqrt{25-d\over3}=Q_L$) and a spin-2 $bc$ system
($c_{gh}=-26$). In components, the DDK realization
of the $N=2$ twisted topological algebra reads \cite{BeSe2} \cite{BeSe3}

\BE \cL_m=\hL_m+l_m,\quad
l_m\equiv\sum_{n\in{\oZ}}(m+n):\!b_{m-n}c_n\!:\label{L26}\EE

\BE\cH_m=\sum_{n\in\oZ}:\!b_{m-n}c_n\!:{}-
{}\sqrt{{3-\ctop\over3}}I_m
\,,\qquad \cG_m=b_m\,,  \label{HG}\EE

\BE\cQ_m=2\sum_{p\in\oZ}c_{m-p}\hL_p
+\sum_{p,r\in\oZ}(p-r):\!b_{m-p-r}c_pc_r\!:{}+
{}2\sqrt{{3-\ctop\over3}}m
\sum_{p\in\oZ}c_{m-p}I_p+{\ctop\over3}(m^2-m)c_m~,\label{Q26}\EE

\noi
and the chiral primary states can be written as
$\ket\P = \ket\Psi \otimes c_1 \ket0_{gh}$,
where $\ket\Psi$ is a primary state in the "matter + scalar" sector
(for a spin-2 $bc$ system $c_1 \ket0_{gh}$ is the {\it true} ghost vacuum
annihilated by all the positive modes $b_n$ and $c_n$).

\vskip .2in

In both realizations of the $N=2$ twisted
topological algebra,
 the relation between the matter central
charge and the {\it topological central charge} turns out to be
\cite{BeSe2} \cite{BeSe3}

\BE d={(\ctop+1)(\ctop+6) \over\ctop-3}\label{d(c)} \EE

\noi
that implies $d\leq 1$ or $d\geq 25$ and $c\neq 3$.

Thus we see that the non-critical bosonic string, for $d\leq1$ or
$d\geq25$, provides a realization of the $N=2$ twisted topological
algebra.

\vskip .4in

\centerline{\bf Acknowledgements}
It is a pleasure to thank our collaborator Aliosha Semikhatov with whom
 most of the results presented here were obtained.

\end{document}